\documentclass[12pt]{article}
\usepackage{graphicx}
\begin{document}

{\bf \large Emergence of firms in $(d+1)$-dimensional work space}

\bigskip
\noindent
G Weisbuch$^1$, D Stauffer$^2$, D Mangalagiu$^3$, R Ben-Av$^4$, S Solomon\\ 
Racah Institute of Physics, Hebrew University, IL-91904 Jerusalem, Israel

\bigskip
\noindent
$^1$ Visiting from Laboratoire de Physique Statistique, \footnote{Laboratoire associ{\'e}
au CNRS (UMR 8550), {\`a} l'ENS et aux Univ. Paris 6 et Paris 7}\\
 Ecole Normale Sup\'{e}rieure, F-75231 Paris, France \\ {\tt gerard.weisbuch@gmail.com}

\noindent
$^2$ Visiting from Institute for Theoretical Physics, \\
Cologne University, D-50923 K\"oln, Euroland

\noindent
$^3$ Management and Strategy Department, Reims Management School,\\ 59, rue Pierre
Taittinger, F-51061 Reims Cedex, France

\noindent
$^4$ Department of Software Engineering, Jerusalem College of Engineering (JCE), Israel 

\bigskip
\section{Introduction}

  Standard micro-economics 
concentrate on the description  of markets but is seldom interested in
production. Several economists discussed the concept
of a firm, as opposed to an open labour market where entrepreneurs
would recrute workers on the occasion of each business
opportunity.   Coase \cite{Coase} is one of them, who
explains the existence of firms as institution 
because they reduce the transaction costs with
respect to an open labour market.

  Whatever the rationale proposed by economists
to account for the existence of firms, their
perspective is based on efficiency and cost 
analysis. Little attention is paid to the dynamics of
emergence and evolution of firms.

  The aim of the present manuscript is to check
the global dynamical properties of a very simple model
based on bounded rationality and reinforcement learning.
 Workers and managers
are localised on a lattice and they choose collaborators
on the basis of the success of previous work relations.
The choice algorithm is largely inspired
 from the observation and modeling of 
long term customer/sellers relationships
observed on perishable goods markets discussed in 
  Weisbuch etal\cite{Weisbuch} and   Nadal etal\cite{Nadal}. 

  The model presented here is in no way an alternative to Coase.
We describe the build-up of long term relationships which do reduce 
 transaction costs, and we deduce the  dynamical properties
of networks built from our simple assumptions.

\section{The model}

\subsection{The model}

   Let us imagine a production network of workers:
we use the simplest structure of a lattice: at each node 
is localised a ''worker'' with a given production capacity
of 1. Business opportunities of size $Q$ randomly strike ''entrepreneur''
sites at the surface of the lattice.

  The work load received by the entrepreneur is too
large to be carried out by her: she then then distributes it randomly
to her neighbours upstream; let us say that these neighbours
are her nearest neighbours upstream.
 We postulate two mechanisms here: a probabilistic choice 
process according to preferences to different neighbours,
and the upgrading of preferences by as a function of previous gains.
The probability of choosing neighbour $j$ is given by
the logit function:

\begin{equation}
 p_j = \exp(\beta J_{j})/\sum_{k=1}^{nb} \exp(\beta J_{k})
\end{equation}

  where the sum extends to all neighbours of the node.
The preferences $J_j$ are updated at each time step
according to:

\begin{equation}
 J_{j}(t)= (1-\gamma)J_{j}(t-1) + q_j(t)
\end{equation}

where $q_j(t)$ is the work load attributed to node $j$.

  One time step corresponds to the distribution of work loads
across the set of collaborators of the entrepreneur
who received the work load. 

    A series of work loads strike the entrepreneur
at successive time steps. We want to characterise the
asymptotic structure generated by a large number of 
work loads presented in succession to the entrepreneur.

\subsection{The algorithm}

Workers are placed on a $(d+1)$-dimensional hyper-cubic lattice of height 
$L_z$. Each hyper-plane $line = 1,2, \dots, L_z$ is a lattice of linear 
dimension $L$ with $L^d$ sites and periodic (helical) boundary conditions.
A workload $Q$ is distributed from the top level (hyper-plane) $line=1$
upstream, in steps from level $line$ to level $line+1$, until $Q$ different
workers (sites) $i$ each have a local workload $q_i = 1$. All local workloads 
$q_i$ are integers $1,2, \dots, Q$. 

One iteration corresponds to the downward distribution of one workload and
proceeds as follows: Initially all sites have workload zero. A new workload 
arrives at the central site of the top level.
 Thereafter, each site $i$ on level
$line$ having a local workload $q_i > 1$ distributes the surplus $q_i - 1$ to 
its $n_{nb} = 2d+1$ nearest and next-nearest neighbours $j$ on the lower level
$line+1$, in unit packets $q_i \rightarrow q_i-1$. For this purpose it selects,
again and again, randomly one such neighbour $j$ and transfers to it with 
probability $p$, given by equation (1) 
one unit of workload, increasing by one unit the preference
 $J_{ij}$ 
storing the history of work relations. After site $i$ has distributed its
workload in this way to the lower level of hierarchy and has only a remaining
unit workload, the algorithm moves to the
next site having a local workload bigger than unity. The whole iteration stops 
when the lowest level $line = L_z$ is reached or when no site has a local 
workload above unity. Then all workloads $q_i$ 
are set back to zero, all stored preferences $J$ are diminished by a factor 
$1 - \gamma$, and a new iteration 
starts, influenced by the past history stored in the preferences $J_{ij}$.

\section{Simulation results}

\subsection{Equilibration}

Fig.1 shows that for the chosen parameters some stationary equilibrium is 
obtained between the increase of sum of all $J_{ij}$, called the flow,
 due to new work, and the decrease
of the flow due to the forgetting parameter.
 The depth of the load pattern in the
lattice also increases and finally reach saturation as the flow.
These dynamics are similar in lower (1+1) and higher (1+4) dimensions.

\subsection{Snakes and blobs}

According to $\beta$, $\gamma$ and load values, after many iterations
two dynamical regimes 
are observed: a quasi-deterministic regime such that only one link out of
$2d+1$ is systematically chosen 
resulting in a ''snake'' portion of the work pattern,
 and a random regime 
where all 3 links are used, resulting in a ''blob''
 portion of the work pattern.
The interface between the two regimes corresponds to
\begin{equation}
  \beta (q(z)-1)/\gamma = {\rm Constant} 
\end{equation}
 where $q(z)-1$ is the work load distributed by a node
of charge $q(z)$ at depth $z$.
 Because the work load to be distributed, $q(z)-1$ decreases 
with increasing depth $z$, there is a given depth where the interface
 between the deterministic regime and the random regime
is located.

 Figure 2 displays workloads obtained 
 in a (1+1)-dimensional lattice. On this figure the snakes
 extends from the initial load of
20 to the load of 7 followed by a small blob of height 3. Parameters for this
simulation were $\gamma=0.3, \beta=0.3$.

 A mean field theory, proposed in a different context by
 Weisbuch etal\cite{Weisbuch} and   Nadal etal\cite{Nadal}
, predicts a transition
between the head and tails regime at a depth $z$ such that:

\begin{equation}
  \frac {\beta * (q(z)-1)}{\gamma}= 2d+1 
\end{equation}

  For larger values of $q(z)$, all the work load is transfered 
to a single neighbour with a preference coefficient of  $(q(z)-1)/\gamma$,
and all other coefficients are 0. For lower values of $q(z)$,
all preference coefficients are small, with possible fluctuations
around the interface. These predictions are verified in  figure 3
computed for simulations in 1+1,
1+2 and 1+3 dimensions.

\begin{figure}[hbt]
\begin{center}
\includegraphics[angle=-90,scale=0.4]{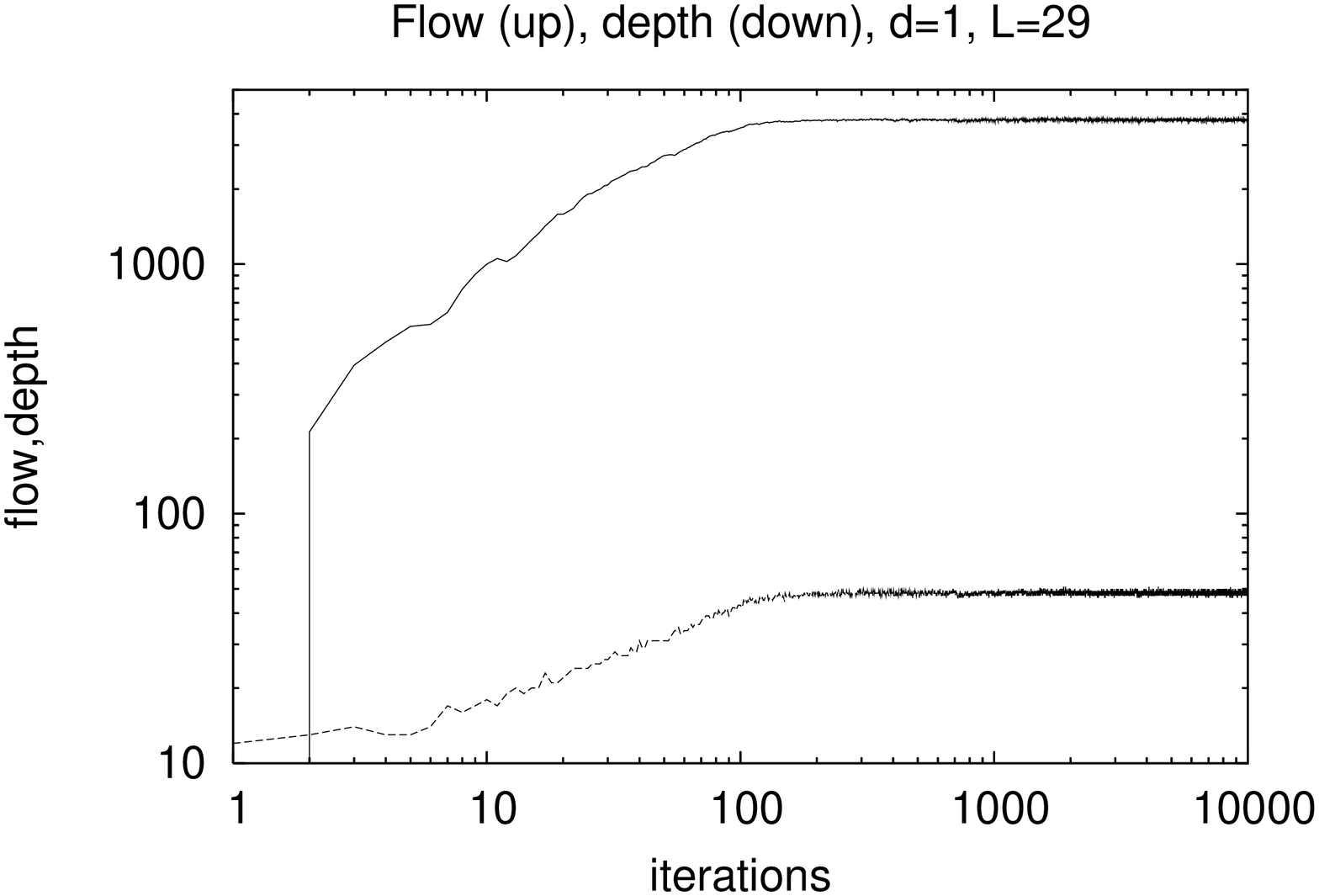}
\includegraphics[angle=-90,scale=0.4]{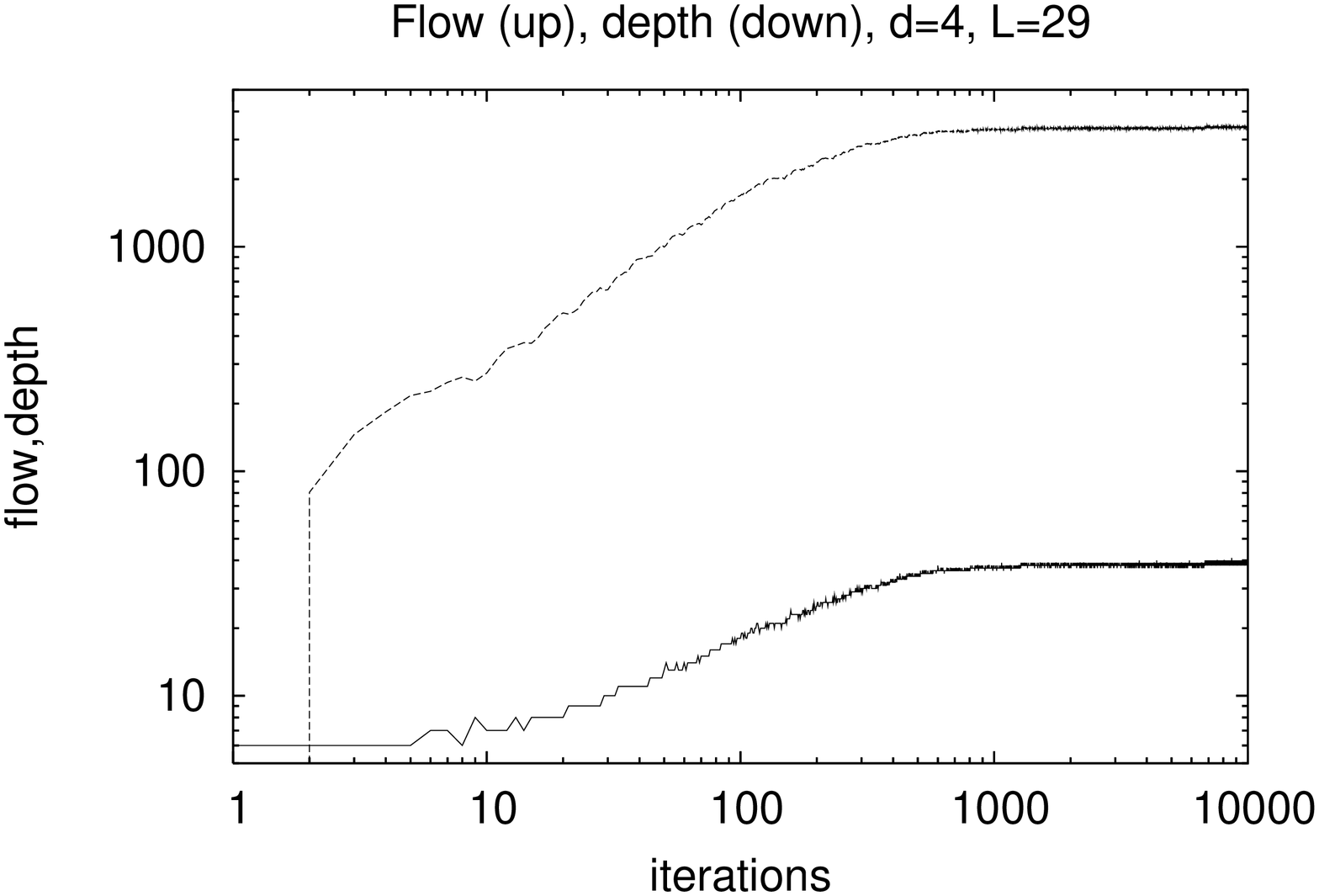}
\end{center}
\caption{Top: Equilibration of the sum of all preference coefficients
 (top curve),
and depth  at $\beta = 0.1, \; \gamma=0.3, \; 1+1 {\rm dimensions}, Q=60$, averaged over 
10,000 iterations.  Bottom: Same parameters except for $\ d=4$.
}
\end{figure}

\begin{figure}[hbt]
\begin{center}
\includegraphics [scale=0.9]{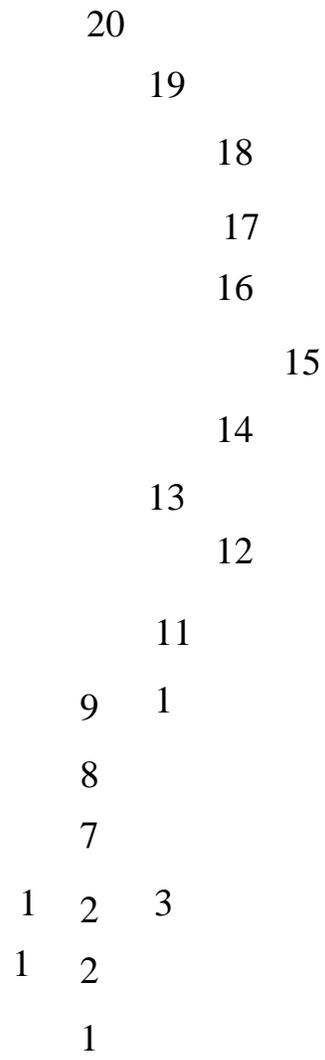}
\end{center}
\caption{  One instance of the work load repartition  from an initial load of 20
at the top site until the lower line. The load is initially distributed with a strong
  preference for one neighbour out of three and is then more uniformly from load 7.
$\beta=0.3, \gamma=0.3, Q=20$.
}
\end{figure}

\begin{figure}[hbt]
\begin{center}
\includegraphics [scale=0.3,angle=-90]{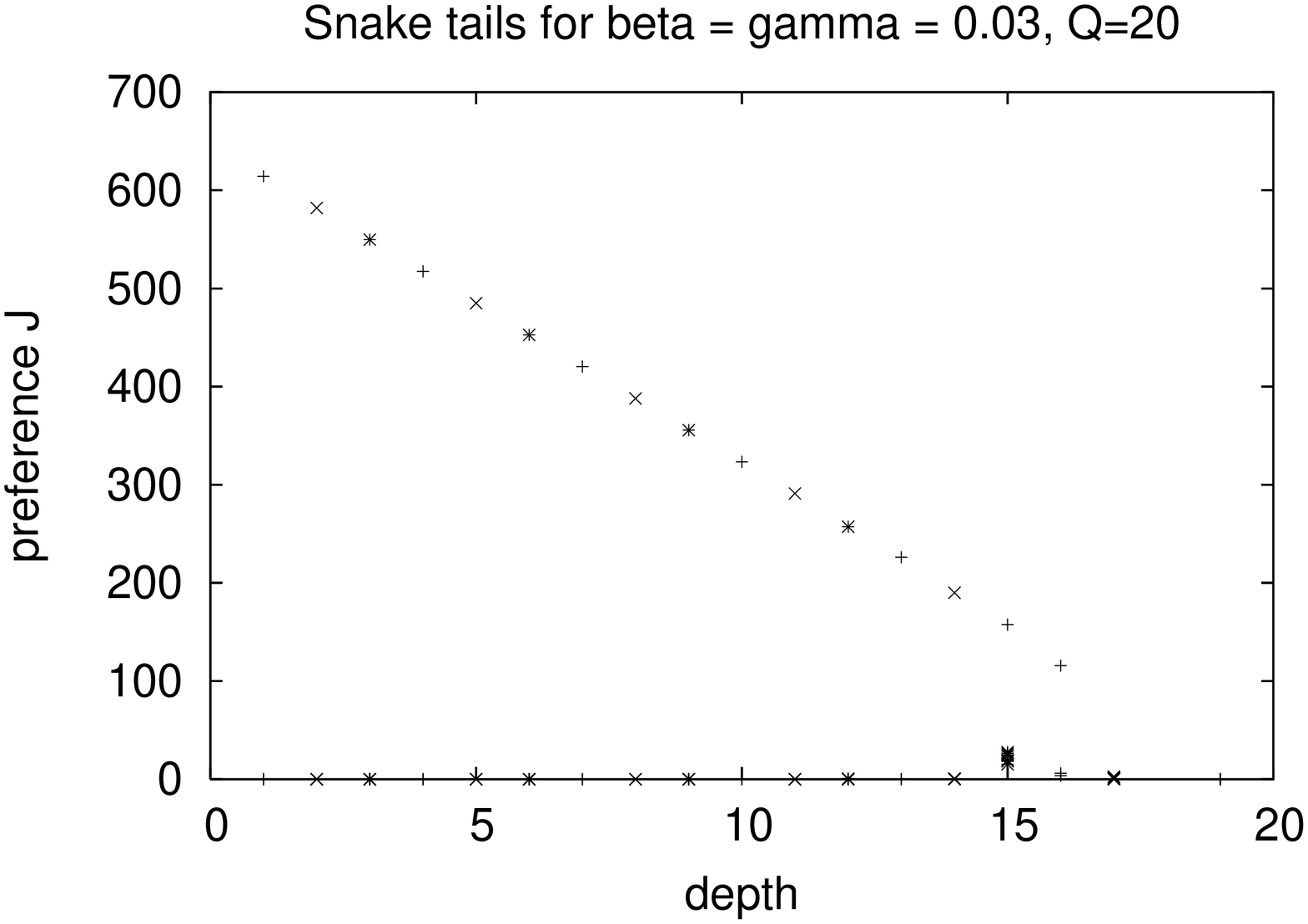}
\includegraphics [scale=0.3,angle=-90]{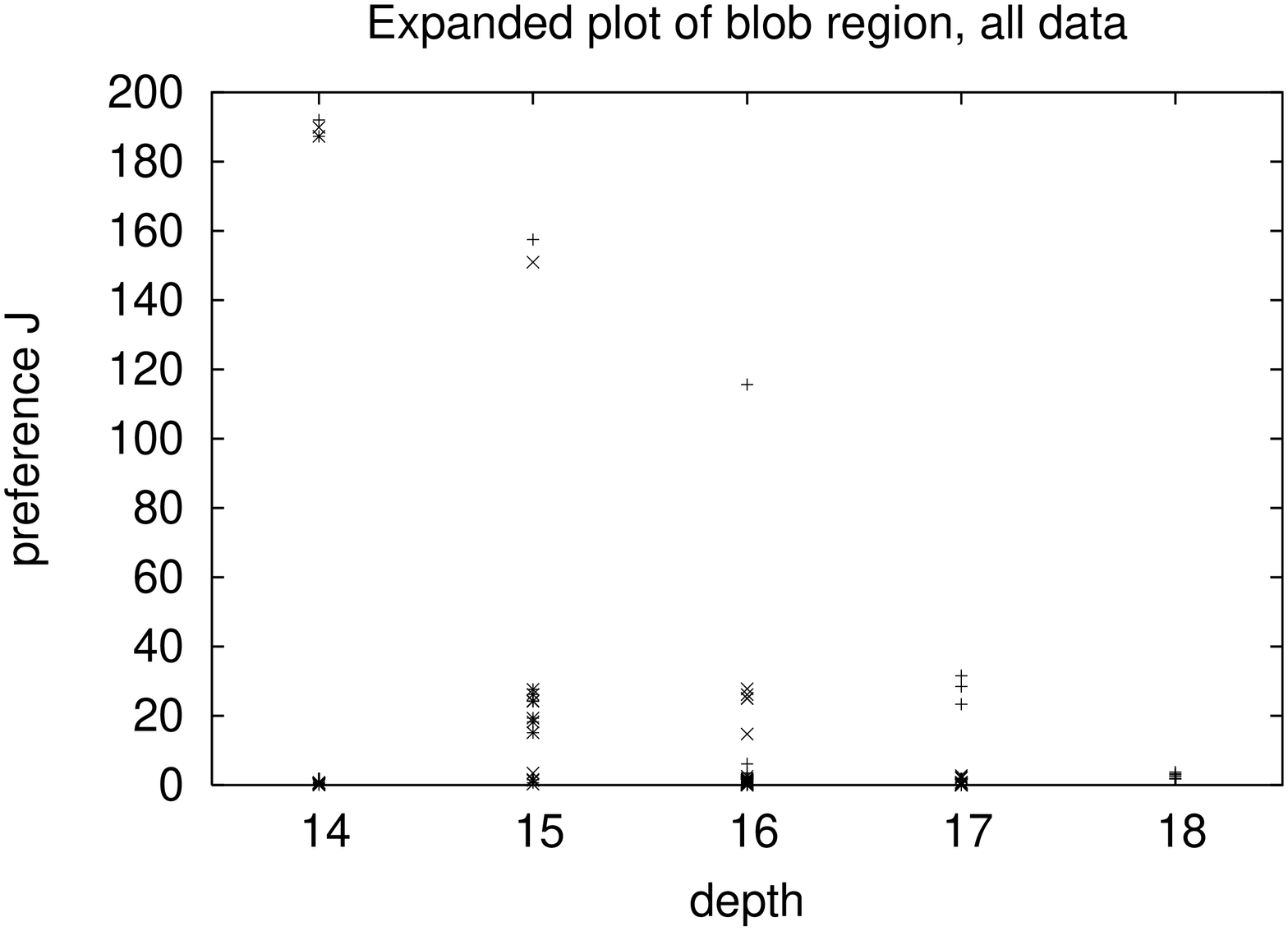}
\end{center}
\vspace{2cm}
\caption{ Evolution in $1+d$ dimensions of the preference coefficient with distance from the ''surface''.
For smaller depths, in the tail region, the preference coefficients are either
 strong ($(q(z)-1)/\gamma$) and independent of the dimension $d$, or zero. A transition is observed
around resp. charges of 5, 4 and 3 rather than for $2d+1=$ 7, 5 and 3 resp. as predicted by the 
mean field theory (equation (3). Top part: Overall picture emphasising the snake.
 Bottom part: Enlargement of blob region. $Q = 20, \; \beta = 0.03, \; \gamma=0.03, \; d=1 (+), 2 (x), 3 (*) $.
}
\end{figure}

\begin{figure}[hbt]
\begin{center}
\includegraphics[angle=-90,scale=0.35]{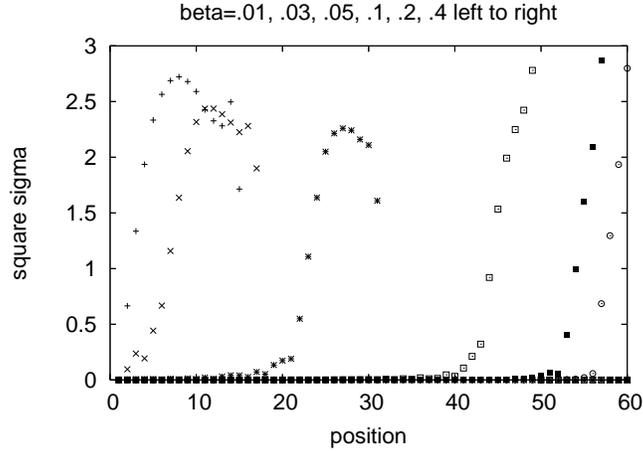}
\end{center}
\caption{Variation with vertical position of the square width of the working 
region within a horizontal hyper-plane (= line), averaged over the last 5,000
of 10,000 iterations.  $ \gamma=0.3, \; d=1, Q=60$
 Similar results were obtained for $d=2$ and 3.
}
\end{figure}

\begin{figure}[hbt]
\begin{center}
\includegraphics[angle=-90,scale=0.35]{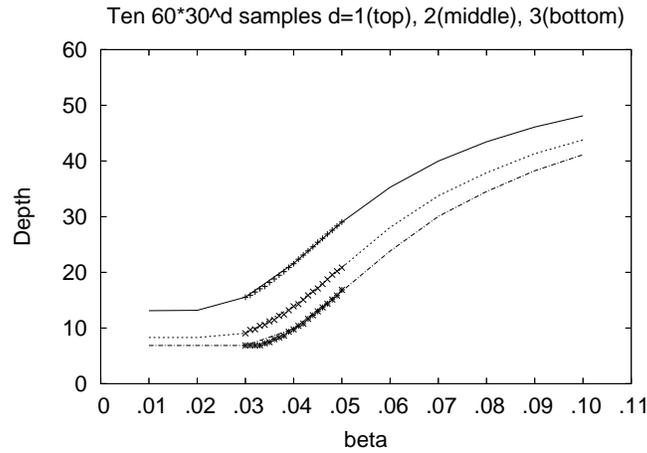}
\end{center}
\caption{Variation with $\beta$ of the vertical position of the lowest 
working plane for several dimensions (1+1,1+2,1+3).
 Enlargement to $L=300$ or $L_z=600$ gives no significant changes.
}
\end{figure}

\begin{figure}[hbt]
\begin{center}
\includegraphics[angle=-90,scale=0.5]{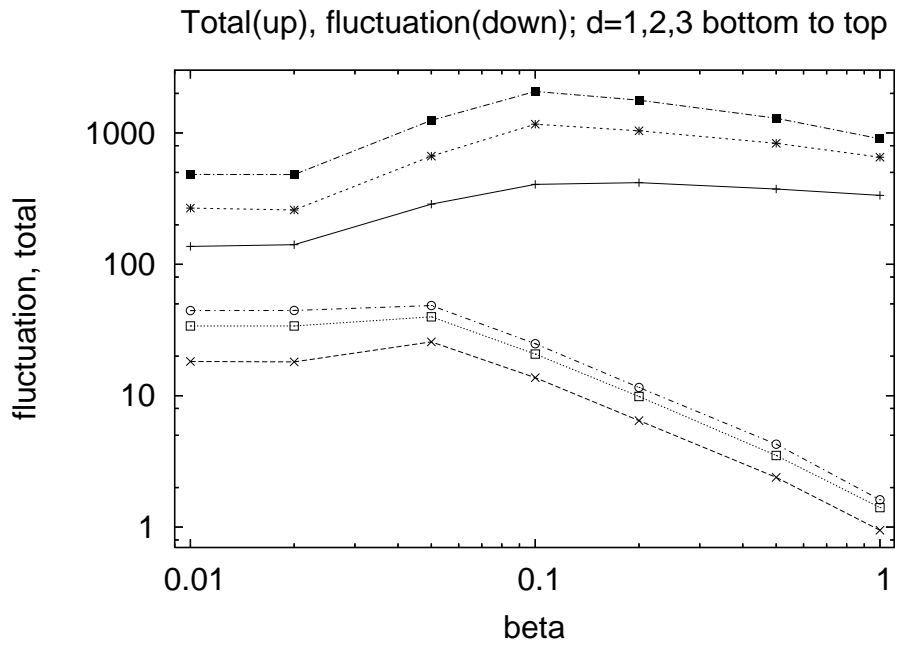}
\end{center}
\caption{Variation with $\beta$ of the total number of different people who 
worked during at least one of the 10,000 iteration (three top curves), and of 
the fluctuations in the work force (three bottom curves); $L=30, \; L_z = 60$,
one sample. Date are averages over the last 1000 iterations.
}
\end{figure}

\begin{figure}[hbt]
\begin{center}
\includegraphics[angle=-90,scale=0.3]{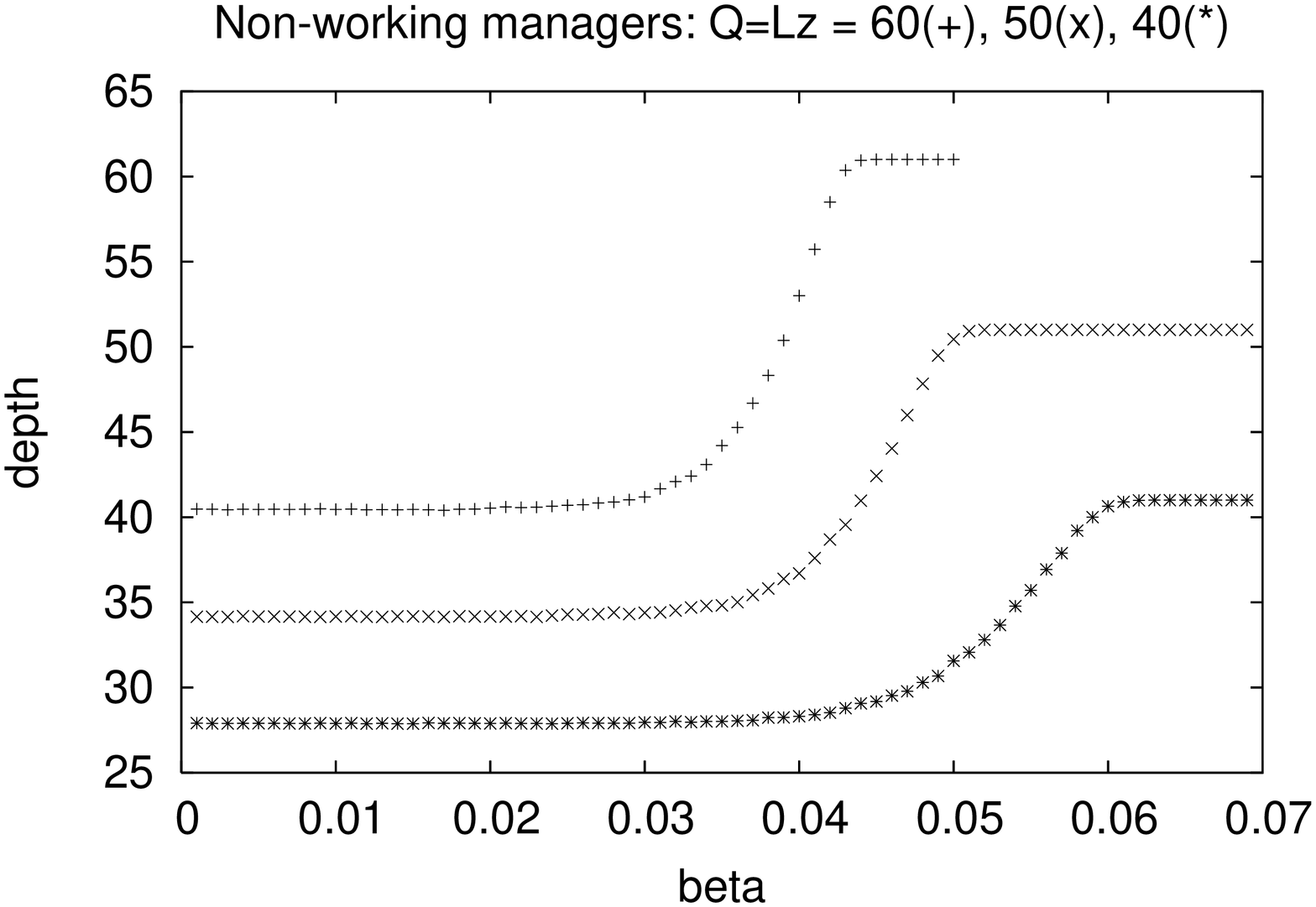}
\includegraphics[angle=-90,scale=0.3]{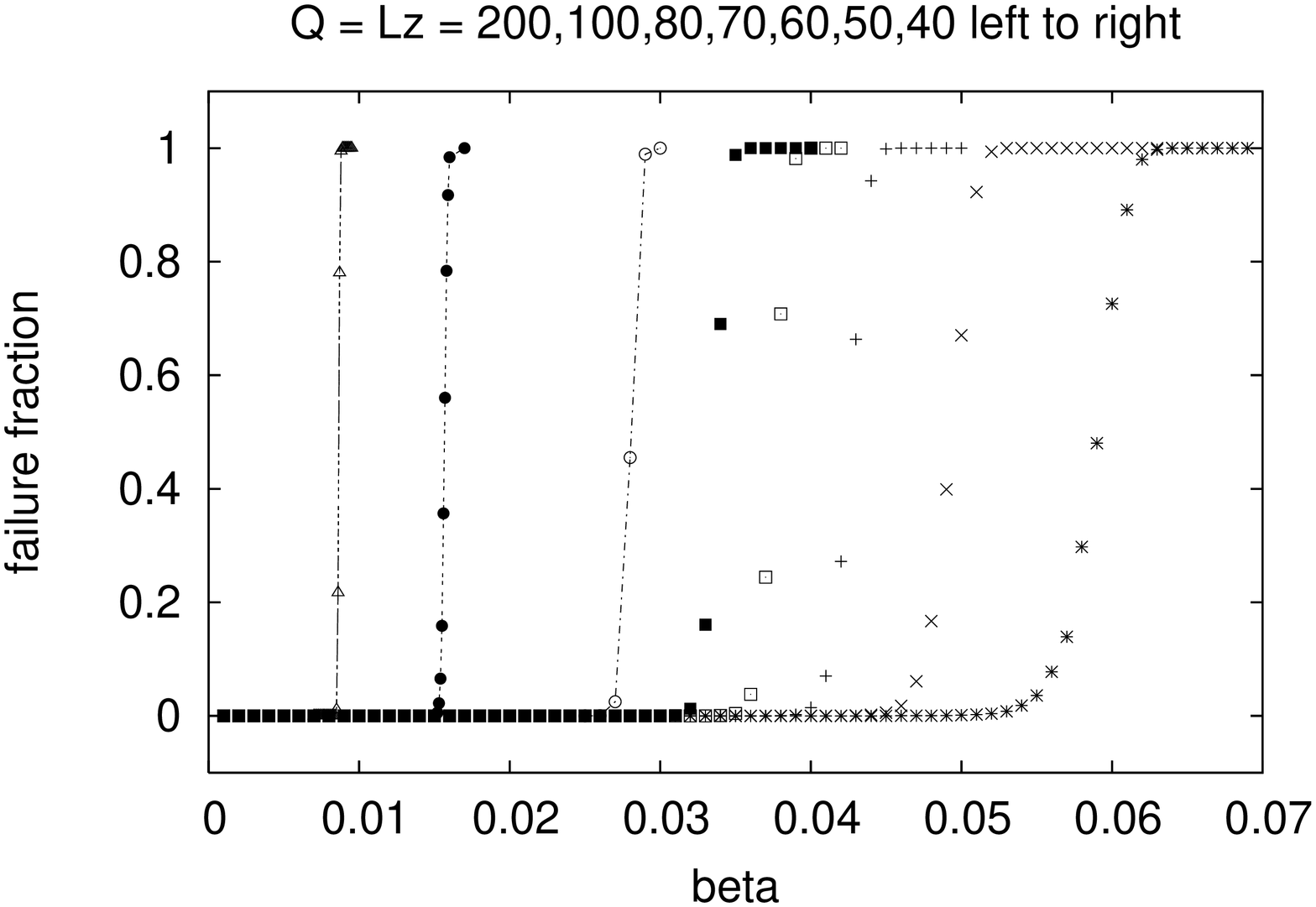}
\includegraphics[angle=-90,scale=0.3]{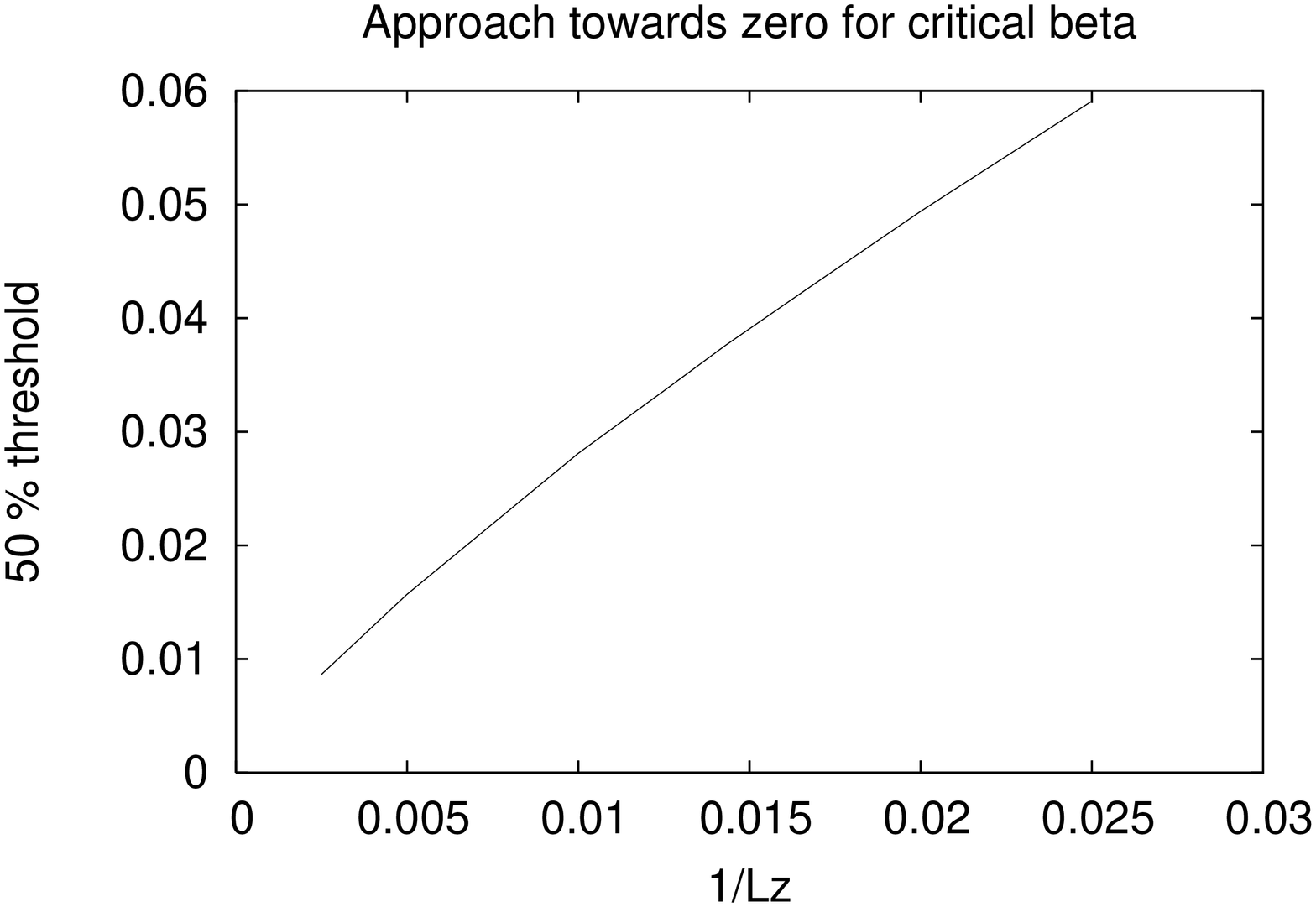}
\end{center}
\caption{Results for the non-working managers model;
 ten samples for 10,000 iterations
each, $d=1, \; L = 30, \; Q =L_z$ increases from right to left.
 Top: Average 
position of lowest working plane. Middle: Average fraction of failures where
process hits the lattice bottom at $L_z$.
 Bottom: $\beta$ value according to middle part 
where the failure fraction reaches 50 percent.$\gamma=0.3$.
}
\end{figure}

\begin{figure}[hbt]
\begin{center}
\includegraphics[angle=-90,scale=0.6]{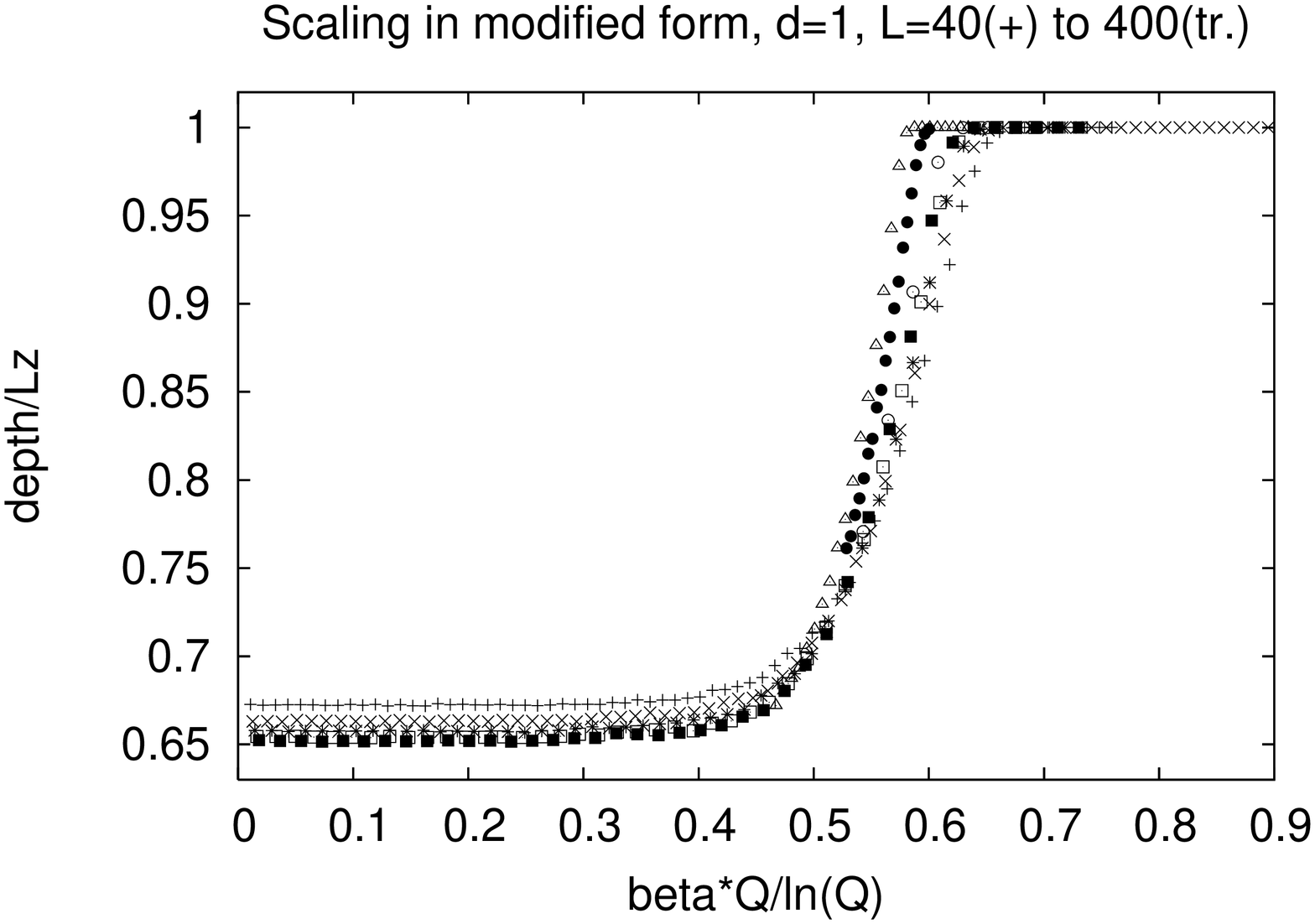}
\end{center}
\caption{Scaling plot of figure 7, top, with approximate collapse
for different $Q=Lz$ between 40 to 400.
}
\end{figure}

Figure 4 is a more systematic test in 1+1 dimension of this dynamics.
 We here plotted in the mean square distance of the 
positions of the workers in each hyper-plane = line from the position of the
highest worker concentration (more precisely:
 from the center of mass of their
distribution). In the tail this squared width is exactly zero (left part),
while in the blob (right part for each curve) it has a peak. The peak
position shifts from small depths (close top plane, no snake) to large depths
(close to bottom plane at 60, long snake), when beta increases. Similar plots
of the snake lengths were obtained for $d=2$ and 3 (not shown); also one test
for $d=4, \; L = 29$ displays an interface between zero and positive 
width.

We plot in Fig.5 the average (over ten samples of
10,000 iterations each) of the position of the lowest hyper-plane touched by
the work distribution process as a function of $\beta$. Although the 
statistics obtained are a clear indication that the 
depth of the system follows the same trend from 1+1 to 1+3 dimensions
they are not directly interpreted. 
  The measured depth is in fact the sum of the length of the snake part
plus the height of the blob part. Both parts vary with $\beta$.
The snake length is obtained from equation (3) since $q(z)$
is simply $Q - z$. But the blob height depends upon the charge 
at the interface
and we don't have any simple analytic expression for it. 

 Figure 6 shows the total number $N_t$ of sites which
were involved in at least one of the 10,000 iterations, i.e. the total
work force with long-time employment, and also at the fluctuations $N_f$ in
the work force from one iteration to the next. (Thus $N_f$ is the number of
sites which are used at iteration $t$ and not at time $t-1$, or the other way
around.) We see that for large $\beta$ the fluctuations are diminished (the same
snake tail again and again passes on the work), but this
decrease is accompanied by an increase in $N_t$, an effect which helps the 
labour market but not the company. 

In the above version, also the managers who distributed work to their lower
neighbours took over one work unit each for themselves. If instead they 
give on all the workload given to them (provided it was at least two units),
then each iteration requires not only $Q$ people as above, but $Q$ workers plus
a fluctuating number of managers. Moreover, if the snake hits the bottom
line at depth $L_z = Q$, part of the work is never finished. This is hardly 
an efficient way to run a business, but Figure 7 and 8 show a much sharper transition,
from a localised cluster at low $\beta$ to delocalized snake tails at high
$\beta$. 

\section{Conclusions}

   The simple reinforcement learning presented here
does end-up in a metastable path in the worker space
represented here by the snake + blob picture, which we interpret as a firm.
On the other hand we would rather imagine firms as hierarchical
structures such as trees \cite{Toulouse,hiera}
. Because of the blob-snake sharp transition as a function of $z$,
we never observe a well balanced tree with a selection at each node
of several preferred collaborators, but rather either a nearly complete preference
for one neighbour or roughly equal preference for all.

  In conclusion, the present model explains the metastability
of employment relations in the firm, but something has to be added to it
to explain the more efficient workload repartition observed in real firms.

  The present manuscript was written during GW and DS stay at the physics
department of the Hebrew University in Jerusalem which we thank for its
hospitality. It was supported by GIACS, a 
Coordination Action of the European communities.

\end{document}